# Reconstruction of the deep history of "Parent-Daughter" relationships among vertebrate paralogs

Haiming Tang*, Angela Wilkins
Mercury Data Science, Houston, TX, 77098
* Corresponding author

**Abstract:**

Gene duplication is a major mechanism through which new genetic material is generated. Although numerous methods have been developed to differentiate the ortholog and paralogs, very few differentiate the "Parent-Daughter" relationship among paralogous pairs. As coined by the Mira et al, we refer the "Parent" copy as the paralogous copy that stays at the original genomic position of the "original copy" before the duplication event, while the "Daughter" copy occupies a new genomic locus.

Here we present a novel method which combines the phylogenetic reconstruction of duplications at different evolutionary periods and the synteny evidence collected from the preserved homologous gene orders. We reconstructed for the first time a deep evolutionary history of "Parent-Daughter" relationships among genes that were descendants from 2 rounds of whole genome duplications (2R WGDs) at early vertebrates and were further duplicated in later ceancestors like early Mammalia and early Primates.

Our analysis reveals that the "Parent" copy has significantly fewer accumulated mutations compared with the "Daughter" copy since their divergence after the duplication event. More strikingly, we found that the "Parent" copy in a duplication event continues to be the "Parent" of the younger successive duplication events which lead to "grand-daughters".

**Data availability:** we have made the "Parent-Daughter" relationships publicly available at https://github.com/haimingt/Parent-Daughter-In-Paralogs/

## Introduction

Gene duplication has been widely accepted as a shaping force in evolution (Zhang, et al. 2001; Zhang 2017). It provides genetic redundancy, releases the selective pressure, and possibly yields new functions. There have been many models of functional changes after gene duplications (Innan and Kondrashov 2010). The 2 most famous models are the neo-functionalization model (Teshima and Innan 2008), where one copy accumulates mutations and develops a new function and the other copy serves as a "spare part" and retains the original function and the sub-functionalization model where functions of the original gene are distributed on the two copies while they both accumulate mutations (Rastogi and Liberles 2005). Although numerous methods have been developed to identify orthologs and paralogs (Thomas 2010), very few have attempted to discern the "Parent-Daughter" relationship among duplicated gene pairs (Han and Hahn 2009). The concept of "parent-daughter" relationship among paralogs is first described in 2007 (Jiang, et al. 2007) and later coined by Mira Han et al in 2009 (Han and Hahn 2009). After a gene duplication event, one or more extra copies of the genes are added the genome, the "parent" copy stays at the same location compared with the original copy before the duplication event and the "daughter" copy is a replicate that occupies a new locus.

Previous method for deciphering the "Parent-Daughter" relationship is based on the assumption that the "parent" copy preserves the longer synteny with the outgroup gene than the daughter copy. However, while checking the complex evolutionary histories of gene duplication in numerous phylogenetic trees, we found that "parent-daughter" relationships of paralogs should be examined with stricter criteria. By combining the idea of synteny evidence comparison and the reconstruction of ancestral gene duplications, we build deep evolutionary history of "parent-daughter" relationships in successive duplications in the vertebrate clade.

The famous 2R hypothesis was first proposed by Susumu Ohno in 1970 (Ohno 1970). It states that the two whole genome duplications had shaped the genome of early vertebrates. The 2 rounds of whole genome duplications yielded 4 copies of each chromosome, which after extensive gene losses and genomic recombination left 4:1 ratio syntenic regions with conserved both gene content and gene orders in extant vertebrate species. In our previous study (Tang and Thomas 2018; Tang, et al. 2018), we have reconstructed the duplication events in early vertebrates from 17 extant species, and summarized the syntenic evidences to the ancestral early vertebrate genes. We examined syntenic evidences from both within and between genomic regions using the extracted vertebrate genes that are descendants of the inferred duplication events in early vertebrates. We have found that more than 55% of the duplication events at early vertebrates have synteny evidences, indicating that they are from large segmental duplications instead of tandem duplications.

While checking phylogenetic trees with duplication events at ceancestors younger than early vertebrates, like "Mammalia" and "Primates", we found deciphering of "Parent-Daughter" relationships in multiple duplication events to be complicated. Duplications at these later periods could copy the "parent" copy or the "daughter" copy in older previous duplication events. While using the simple assumption that, "parent" copy preserves the longer synteny with the outgroup gene than the daughter copy, we could only determine one pair of "Parent-Daughter" relationship, instead of a deep history of "Parent-Daughter" relationships at successive duplication events.

Here we present a novel method to decipher the "parent-daughter" relationship by combining evidences from syntenic regions with ages of inferred duplication periods. The idea is that whole genome duplications at early vertebrates formed syntenic regions with preserved contents and gene orders, thus the synteny evidence examined from extant vertebrates may show ancient "daughter" gene copies from the early vertebrate duplications. These daughter gene copies could be further duplicated and become the "parent" copy in later younger duplication events. On the other side, the duplicated copies from an earlier evolutionary period must be "older" than copies from a later period. The young ceancestors do not experience whole genome duplications and if paralogs descedanted from these ceancestors show syntenic evidence, we could infer these paralogs to have stemmed from early vertebrate ceancestor and become the "parent" which gives rise to new copies in later ceancestors. In the case of multiple sequential duplications, we can identify the gene copy that has synteny evidence as the "parent" or "grand-parent" copy from the whole genome duplication events at early vertebrates. Thus, combined with the phylogenetic tree structure and the syntenic evidence, we could enable a deep deciphering of parent and daughter relationships in each of the consecutive duplication event.

We have discovered several different scenarios of the consecutive duplication events in vertebrates. The most common scenario is that only 1 duplication event happened at a younger period after early vertebrate. The other 2 scenarios are that at least 2 consecutive duplication events at various younger periods have happened. For example, an ancestral gene copy of early vertebrates is duplicated at Tetrapoda ceancestors, yielding 2 additional gene copies. Then one of this gene copy at Tetrapoda is further duplicated at Amniota ceancestor, yielding 2 additional gene copies. These 2 scenarios differ as to whether the "Parent" copy or the "Daughter" copy at Tetrapoda ceancestor is further duplicated at Amniota ceancestor.

From a total of 2009 gene families with reconstructed duplications at early vertebrates, we have examined a total of 4225 cases of parent-daughter relationships. We compared the selection pressure of parent and daughter copies by summarizing the branch length from the divergence of these copies until tree leaves. We have found significantly smaller accumulated branch length in the parent copy than the daughter copy, suggesting less accumulated amino acids in parent copy and that the parent copy is significantly more conservative than the daughter copy. Besides, we have found a scenario where "parent" copy in an older duplication event continues to be the "parent" in following duplication events of a younger ceancestor. This scenario is 52.8% more common than the other scenario where the "daughter" copy in the previous duplication event is choice of "parent" copy in the later duplications. Indicating that "parent" copy in a previous duplication event is more likely to stay at its original genomic loci, and give rise to new duplications in successive duplication events.

## Materials and Methods

**Phylogenetic trees and Ancestral reconstructions**

PANTHER is a large collection of protein families that have been subdivided into functionally related subfamilies (Mi, et al. 2017). Hidden Markov models (HMMs) are built for each family and subfamily for classifying additional protein sequences (Eddy 1996). The PANTHER Classifications are the results of human curation as well as sophisticated bioinformatics algorithms (Thomas 2010; Bansal, et al. 2012). In summary, trees of gene families are constructed using 1) an assumed reference species tree, 2) knowledge of all recognizable members of a given family in each genome, 3) protein sequence data for each member, and 4) identification of potentially problematic gene predictions. A gene tree is constructed stepwise by creating "orthologous subtrees" that are connected via inferred duplication and transfer events. Each subtree contains orthologous sequences related by speciation events with at most one gene from each organism, and the tree topology is determined by the known species tree; a duplication event is inferred only when there is genomic proof that a duplication occurred, viz. when, during the iterative process, a given subtree contains more than one gene from the same species (within-species paralogs). The tree is reconciled with the species tree, meaning that each internal node is labeled with the event type (speciation, duplication, transfer): speciation nodes are labeled with the cenancestor from the species tree, and duplication nodes are labeled by the oldest ceancestors in descendant speciation nodes (Tang, et al. 2018). In this analysis, the PANTHER database version 10 was used (release date July 2016) which contains more than 1 million genes from 104 genomes. The phylogenetic trees structures that were utilized for this study are summarized in Supplemental Material Part 1.

**Synteny evidence examination**
Synteny represents the conservation of blocks of order within two sets of chromosomes that are being compared with each other (Kasahara 2007). In our previous study, we have examined the synteny evidence of early vertebrate duplication events by summarizing both within and between genomic synteny evidences. Detailed methods could be found in our previous paper (Tang and Thomas 2018). In brief, the genes of 17 extant vertebrate species are extracted from the PANTHER database where they are descendants of the reconstructed duplication events at early vertebrates. Then the extracted genes are re-grouped into chromosomes and scaffolds based on their genomic locations, and the homology relationship among genes are labeled by the duplication node at early vertebrates. Both within and among genomic syntenies were detected using i-ADHoRE 3.0 (Proost, et al. 2012). The synteny evidences on extant genes are then summarized through the phylogenetic tree structure to the corresponding ancestral speciation nodes at early vertebrates. The synteny evidences are summarized in Supplemental Material Part 2.

**Determination of Parent-Daughter relationship among vertebrate paralogs**

We determine the parent-daughter relationship among vertebrate paralogs based on the assumptions that 1) the hypothesized 2 rounds of whole genome duplications at early vertebrates result in syntenic regions that could still be detected from extant vertebrate genomes; and that 2) the duplications that happened after the early vertebrate's period were not from the whole genome duplications, and these relatively new daughter paralogs don't share synteny with parent paralogs. By combining the synteny evidences and the reconstructed duplication time periods of paralogs, we can decipher parent-daughter relationships in situations with consecutive duplications.

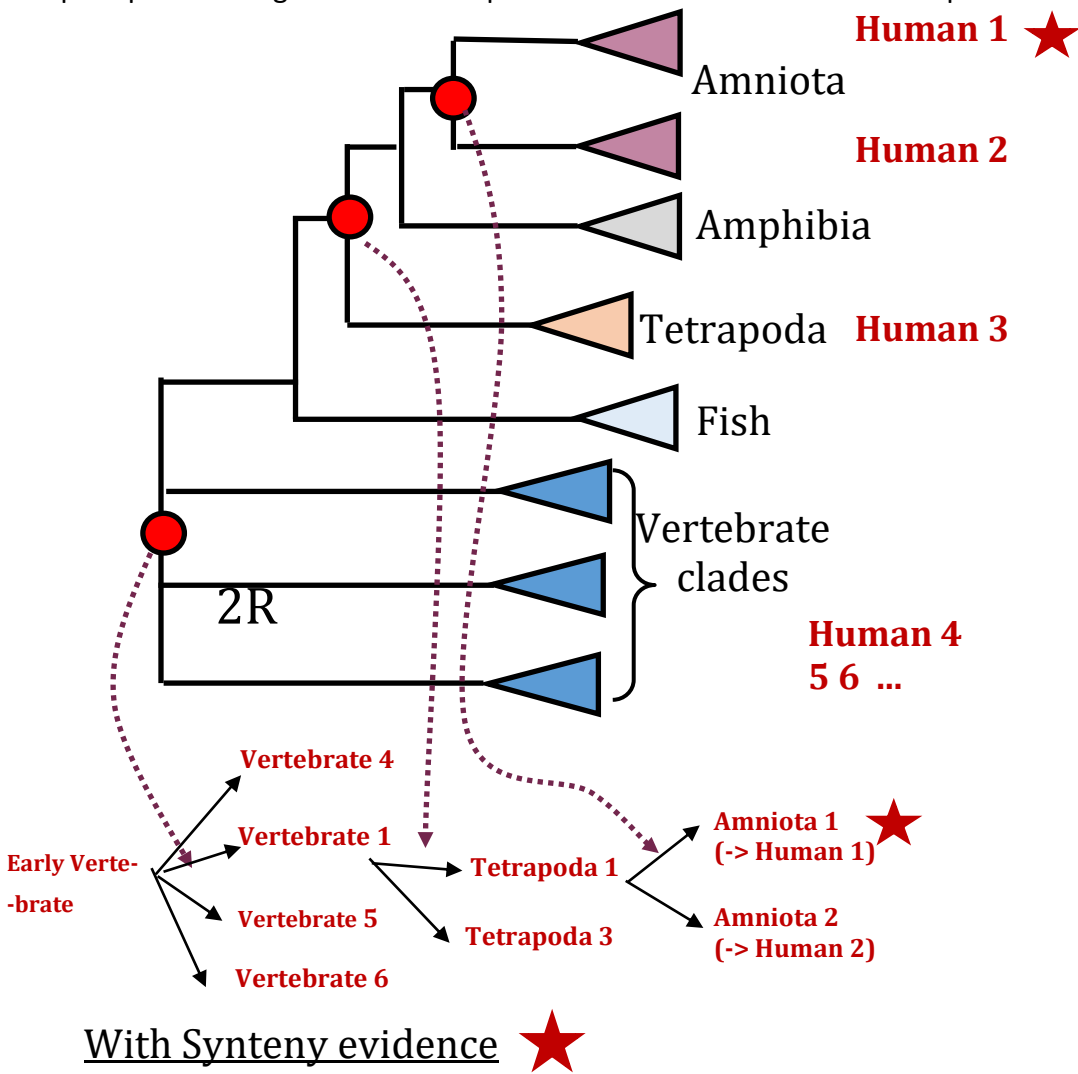

**Figure 1. Determination of Parent-Daughter relationship among vertebrate paralogs**

The figure above shows a hypothesized phylogenetic tree structure, where the red dots represent duplication events and the triangles represent a collapsed clade. For example, the magenta triangle on top represents a monophyletic clade of Amniotes. The "Human 1, 2 …" illustrates the extant human genes in different clades.  The simplified tree structure on the bottom represent the reconstructed evolutionary history of "Parent-Daughter" relationships at each duplication event. The "Vertebrate 4" represents a hypothesized gene in early vertebrates which would be inherited to "Human 4" or "Mouse 4" via speciation events; the "Tetrapoda 1" represents a hypothesized gene in the Tetrapoda ceancestor, which gave rise to an ancestral gene in Amniota ceancestor: "Amniota 1" and finally an extant human gene: "Human 1". The red star indicates that synteny evidence. In this example, the "Human 1" gene has synteny evidence that indicates its origin from the 2 rounds of whole genome duplications at early vertebrates. By assuming the correctness of the reconstructed phylogeny tree structure, we could infer the "Parent-Daughter" relationship of the "Human 1" in each of the duplication events it has gone through. The reconstructed evolutionary history is that during the 2 rounds of whole genome duplications at early vertebrates, 4 copies of genes were created: "Vertebrate 1", "Vertebrate 4", "Vertebrate 5" and "Vertebrate 6". Then "Vertebrate 1" was further duplicated in Tetrapoda ceancestor, giving rise to "Tetrapoda 1" and "Tetrapoda 3". The "Tetrapoda 1" was further duplicated in Amniota ceancestor, adding an additional copy of "Amniota 2" to the genome of the ceancestor Aminota. The "Aminota 1" and "Aminota 2" genes gave rise to "Human 1" and "Human 2" genes via speciation.

We first select protein families that have the inferred duplication events at early vertebrates, and further select the families that have duplications at later evolutionary periods within the vertebrate clade: such as Tetrapoda and Amniota. Such a protein family is illustrated in Figure 1, where the red dots represent duplication events, and the triangles represent monophyletic clades. There are multiple human paralogs in this family: Human 1, Human 2, … We then determine the evolutionary history of paralog pairs by mapping the genes to the tree structure. For example, extant gene Human 1 and Human 2 are in 2 Amniota Clades, generated by a duplication event in Aminota ceancestor (red dot 3). By tracing up, we see a duplication event in Tetrapoda (red dot 2), and an extant Human 3 gene. Based on the phylogenetic tree structure, we could infer that there are 2 possible scenarios in evolutionary history: the first scenario is that a duplication event in Tetrapoda creates 2 copies of gene in ancient Tetrapoda ancestor, one gives rise to "Human 1" in human and the other gives rise to "Human 3", a further duplication event happened in the Amniota ancestor, where the ancestral gene of "Human 2" in Amniota yields "Human 1" and "Human 2". Thus "Human 2" is a "daughter" copy of "Human 2". The second scenario is that "Human 2" and "Human 3" are the 2 copies of genes in the duplication event at Tetrapoda ceancestor, and "Human 2" is the parent copy of "Human 1" in later duplication event at Amniota ceancestor. These 2 scenarios could be differentiated by examining synteny evidences of "Human 1" and "Human 2" with paralogs in other early vertebrate clades including "Human 4", "Human 5" and "Human 6".  2 rounds of whole genome duplications yield 4 chromosome regions with conserved gene contents and orders. 600 million years of evolution since the divergence of early vertebrates have tremendously changed the vertebrate genomes with gene gain, gene loss and recombination. Thus, the syntenic regions detected from extant vertebrate genomes are most likely to be gene block remnants of the whole genome duplications, and the genes within these regions are most likely from the whole genome duplication events at early vertebrates instead of later duplications.

Thus if "Human 1" is found to have syntenic evidence, then we could infer scenario 1 to be the most likely evolutionary history: 2 rounds of whole genome duplications yield 4 copies of early vertebrate genes; "early vertebrate gene 1" was duplicated in Tetrapoda ceancestor, thus giving rise to its daughter copy of "Tetrapoda gene 3", while the "early vertebrate gene 1" was inherited as "Tetrapoda gene 1"; "Tetrapoda gene 1" was later duplicated at Amniota ceancestor, giving rise to a daughter copy of "Amniota 2", while the "Tetrapoda gene 1" was inherited as "Aminota 1". "Human 1" was inherited from "Amniota 1" and "Human 2" was inherited from "Amniota 2" via speciation. Thus, we identify 2 "parent-daughter" relationships here: one at "Tetrapoda", "Human 1" is the parent and "Human 3" is the daughter; the other one at "Amniota", "Human 1" is the parent and "Human 2" is the daughter.

**3 different scenarios of "parent-daughter" relationships in vertebrate paralogs**

While processing more than 8000 protein families in PANTHER, we summarized 3 different patterns of "parent-daughter" relationships in further duplications after early vertebrates. They are illustrated in Figure 2.

The first condition is that, in one or more vertebrate clades that were descendants from the 2 rounds of whole genome duplications, only 1 round of duplication has happened at younger ceancestors in the vertebrate clade. (For a duplication node that has more than 2 descendants, multiple duplications may have happened consecutively in a very short period, corresponding to only 1 ceancestor. For simplicity, we treat these duplication nodes as the same as duplication nodes that have only 2 descendants.) In this condition, only 1 pair of parent and daughter relationship could be identified in paralog pairs that are descendants of the clades from the younger duplication: paralog that has synteny evidence (Human 1 in Figure 2 A) is the parent of the paralog that has no synteny evidence (Human 2 in Figure 2 A).

The second condition is that, in one or more vertebrate clades that were descendants from the 2 rounds of whole genome duplications, more than 1 round of duplications happened at 2 or more different ceancestors in the vertebrate clade. The examination of synteny evidence shows that the paralog with synteny evidence is not within the descendant clades of the youngest ceancestor (like Human 3 in Figure 2 B, that is not in the deepest clade in the phylogenetic tree). Thus, we can only assert the parent-daughter relationship in the paralogs of an older duplication node that has synteny evidence in descendants. In the example illustrated in Figure 2 B, the ancestral "Tetrapoda 3" gene that was inherited as "Human 3" was the parent of a duplicated "Tetrapoda" gene that was further duplicated at Amniota ceancestor which gave rise to "Human 1" and "Human 2" genes, as "Human 3" has synteny evidence with "Human 4-6" in other vertebrate clades. We won't be able to assert the parent-daughter relationship in paralogs in younger duplication nodes deeper in the phylogenetic tree structure. In the Figure 2B example, we are unable to assert whether the "parent" is the ancestral Amniota gene that was later inherited as "Human 1" or a gene copy that was later inherited as "Human 2" for the duplication node at Amniota ancestor.

The third condition is similar with the second condition: in one or more vertebrate clades that were descendants from the 2 rounds of whole genome duplications, more than 1 round of duplications happened at 2 or more different ceancestors in the

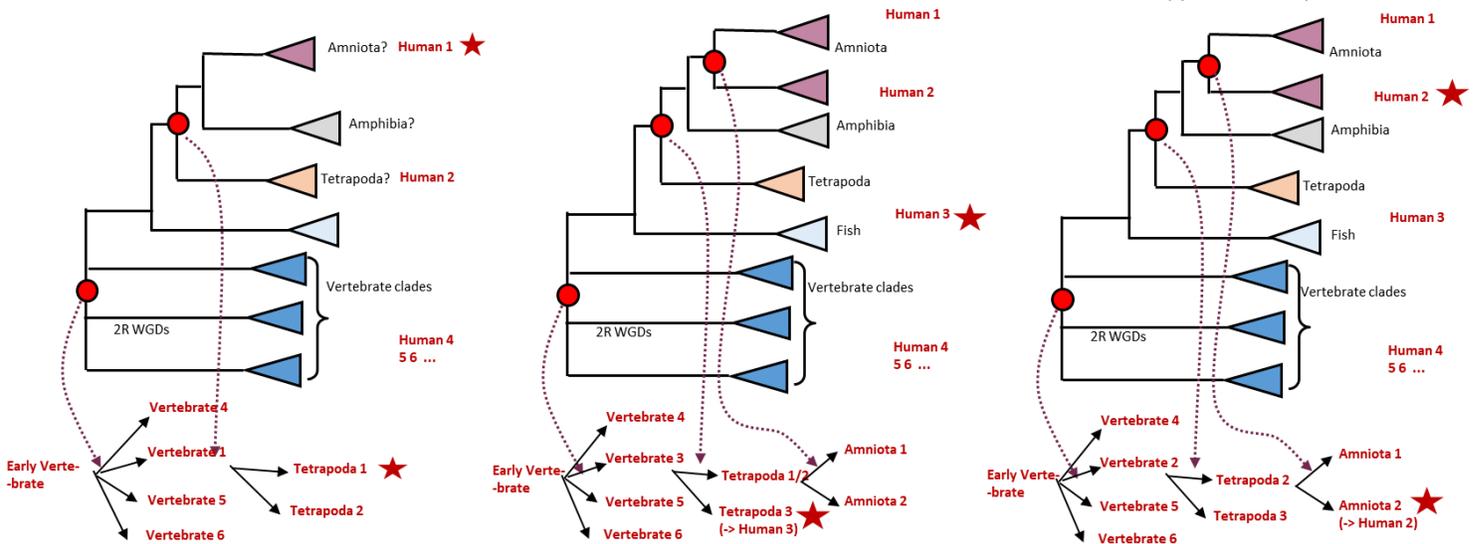

**Figure 2. 3 different scenarios of "parent-daughter" relationships in vertebrate paralogs**
A. Condition 1, vertebrate clade after 2R WGD has 1 and only 1 duplications at younger periods.
B. Condition 2, vertebrate clade after 2R WGDs has more than 1 duplications at younger periods, but the "Parent" copy is not further duplicated. C. Condition 3, vertebrate clades after 2R WGDs has more than 1 successive duplications at younger periods, and the gene with synteny evidence is located within the deepest duplication node.

vertebrate clade. The examination of synteny evidence shows that the paralog with synteny evidence is within the descendant clades of the youngest ceancestor that has duplication. The example in Figure 1 also falls into this condition. The "synteny evidence" with the paralog indicates that the paralog was from the 2 rounds whole genome duplications at early vertebrates and had preserved its loci in the syntenic chromosome regions since then. It's innermost location in the duplication events indicate that the paralog had been the "parent" and gave rise to "daughter" copy in each of the duplication events. In the example show in Figure 2 C, the ancestral "Tetrapoda 1" gene that was later inherited as "Human 1" by speciation was the parent of an ancestral "Tetrapoda 3" gene that was later inherited as "Human 3". In the younger duplication event at Amniota ceancestor, the ancestral "Tetrapoda 1" gene was inherited as

"Amniota 1" gene via speciation and was the parent of a duplicate "Amniota 2" that was inherited as "Human 2".

The scripts to get the 3 different conditions are in Supplemental Material Part 3.

**Comparison of selection pressure between parent and daughter gene copies**

Selection pressure could be measured by protein sequence changes in a specific period (Hahn, et al. 2007). As we want to compare the selection pressure between parent and daughter gene copies, the accumulated protein sequence changes since the divergence of the parent and daughter copies were calculated. Thus, we used the summed branch length from a duplication node to the extant paralogs for analysis.

In condition 1, the calculation of the summed branch length is trivial. The selection pressure for parent is the summed branch length from the duplication node to the paralog with synteny evidence (Human 1 in Figure 2 A), and that for daughter is the summed branch length from the duplication node to the paralog without synteny evidence (Human 2 in Figure 2 A). Condition 2 is trickier as we don't know which paralog in a younger duplication is the parent. Thus, the selection pressure for the "parent" is calculated in the same way as in condition 1 (Human 3 in Figure 2B). But
for the selection pressure of the daughter copy, we get the average of the summed branch length of paralogs that have no clear origin (The average of summed branch length from the duplication node at Tetrapoda ceancestor to Human 1 and Human 2). For condition 3, as multiple "parent-daughter" relationships are identified at different duplication nodes, multiple comparisons are performed for each of the "parent-daughter" relationship. For the example in Figure 2 C, summed branch length from the duplication at Tetrapoda ceancestor to "Human 1" for the parent copy of the duplication event at Tetrapoda and that to "Human 3" for the daughter copy. Besides, the summed branch length from the duplication at Amniota ceancestor to "Human 1" for the parent copy of the duplication event at Amniota and that to "Human 2" for the daughter copy.

A shorter summed branch length indicates less accumulated amino acid changes and indicate more resistance to the selection pressure. Paired T test was used in each condition for significant differences in the branch length. Statistical analysis is done using STATA 13.

## Results

We have analyzed 11928 protein families in PANTHER 10, out of which 2008 (16.8%) have predicted duplication events at early vertebrates and further duplication after early vertebrates. From the synteny evidences of 17 extant vertebrate species in PANTHER, our previous study showed that 55% of the duplication events at early vertebrates still preserve the synteny evidence in their descendants, providing a convincing evidence for the 2R WGDs hypothesis.

By combining the synteny evidences and the reconstructed duplication time periods of paralogs, we can decipher parent-daughter relationships in situations with further duplications in ceancestors younger than early vertebrates. We have summarized the parent and daughter copies and their accumulated branch lengths from the divergence for each different condition in Supplemental Material Part3.

**Table 3. Comparison of the selection pressure for parent and daughter copies**

| Conditions | | Observations | Parent-Daughter copies | Accumulated branch length since the divergence of the parent and daughter copies | | | P |
|---|---|---|---|---|---|---|---|
| | | | | Mean | Std. Dev. | 95% Conf. Interval | |
| 1 | | 1622 | Parent | 0.792 | 0.872 | (0.750, 0.835) | < 0.0001 |
| | | | Daughter | 1.066 | 1.261 | (1.005, 1.128) | |
| 2 | | 576 | Parent | 1.468 | 1.257 | (1.365, 1.571) | < 0.0001 |
| | | | Daughter | 1.667 | 1.452 | (1.548, 1.786) | |
| 3 | From the deepest duplication node | 905 | Parent | 0.816 | 1.092 | (0.745, 0.888) | 0.062 |
| | | | Daughter | 0.859 | 1.164 | (0.783, 0.935) | |
| | From other duplication nodes | 769 | Parent | 1.606 | 1.717 | (1.484, 1.727) | <0.0001 |
| | | 769 | Daughter | 1.376 | 1.247 | (1.287, 1.464) | |

We have identified 3 different scenarios for further duplications after the 2R WGDs. 1662 cases were found for the first scenario where there is only 1 further duplication at younger ceancestor after early vertebrates. The average of the summarized branch length for the parent copy is statistically significantly smaller than the daughter copy (0.792 vs 1.066, p < 0.001), suggesting the "parent" copy is more resistant to amino acid changes. We have identified 576 cases for the second scenario where there are more than 1 further duplications at 2 or more younger ceancestors after early vertebrates. The paralog with synteny evidence was not in the youngest duplication event. This indicates that we could identify the "parent-daughter" relationship for the duplication event that has the paralog with synteny in direct descendants. But the "parent" in this specific duplication event was not the "parent" in successive younger duplications. For condition 2, the average of the summarized branch length for the parent copy is also statistically significantly smaller than the daughter copy (1.468 vs 1.667, p < 0.001), supporting the hypothesis that "parent" gene copy is more resistant to amino acid changes.

Condition 3 is similar with condition 2, but the paralog of synteny evidence is located within the innermost duplication node. This shows that the "parent" copy in an older pervious duplication event continues to be the "parent" in the successive younger duplication events. The fact that there are 57.1% more cases of condition 3 than condition 2 suggests that "parent" copy in a previous duplication event is more likely to continue be the "parent" copy and stay in the original genomic loci if this gene would be successively duplicated in later evolutionary periods. For condition 3, we could identify multiple "parent-daughter" relationships for a single case, thus we collect different sets of summarized branch length for each identified relationship. For the innermost or youngest duplication event, the average of the summarized branch length for the parent copy is marginally smaller than the daughter copy (0.816 vs 0.859 $p < 0.062$), supporting the hypothesis that "parent" gene copy is more resistant to amino acid changes. For the older duplication events than the innermost duplication event, the average of the summarized branch length for the parent copy is also statistically significantly larger than the daughter copy (1.606 vs 1.376, $p < 0.001$). This indicates an interesting scenario where the "Parent" copy has yielded multiple duplicated copies in evolutionary history, but the "daughter" copy that was generated in earlier duplications had accumulated fewer changes compared with the "Parent" copy which has gone through further duplications.

## Discussion

Here we present an unprecedented way to decipher the "parent-daughter" relationship among vertebrate's paralogs that meet strict criteria. There must be further duplication events at younger ceancestors in one or more of the early vertebrate duplication clades. Besides, synteny evidence should be identified in some paralog suggesting it had preserved relative genomic loci during the 600 million of evolution since early vertebrates (Peterson, et al. 2008).

While discussing with other researchers, we found researchers who were opposed to the concept of the "parent" (old) and "daughter" (young) duplicates. One argued that after a duplication event at a ceancestor, all duplicated copies are "new" copies in the new individual or species. we acknowledge this that both copies should have the same "age" and the duplicated copies are identical except their genomic locations. However, this discussion is more philosophical than scientific. Here in this paper, we still use the "parent-daughter relationship" term coined by previous researchers. The "parent" copy refers to the gene copy that stays at the original location as the gene before the duplication event while "daughter" copy is the one that moves to a new genomic position.

The inference of parent-daughter relationship was based on several assumptions: 1) the hypothesis of 2 rounds of whole genome duplications is correct and the homologous genomic regions with preserved gene contents and order in extant vertebrate species are remnants of the 2 rounds whole genome duplications. 2) Gene duplications at ceancestors later than early vertebrates are not from whole genome duplications or large segmental duplications. Thus, synteny evidence does not mistakenly show paralogs that originated from ceancestors later than early vertebrates. 3) The inference of ancestral gene duplications in PANTHER was largely correct. PANTHER first …., then combine together. The inference of ceancestor for a gene duplication is based on the extant species that are present in each of the descendant clades. Considering the possibility of gene losses in specific vertebrates, the inferred ceancestor may be younger than the actual duplication event in evolutionary history.

Few researches have tackled the parent-daughter relationships in the duplication events. Previous research is based on the simple assumption that the parent copy has more syntenic evidences than the daughter copy. This assumption still holds in our research, but we have greatly broadened the examination of parent-daughter relationships by examining specific phylogenetic tree structures. Specially in condition 3, where multiple "parent-daughter" relationships could be inferred from duplication events at successive ceancestors.

We argue several conditions where the parent daughter relationship among paralogs could not be easily discerned. For example, we will make a mistake if there are no further duplications after the hypothesized 2 rounds of whole genome duplications at early vertebrates. By looking at the duplications in early vertebrates alone, a phylogenetic tree could be built via the similarities of the 4 paralogs in each of the early vertebrate clades. Like "Human 1", "Human 4", "Human 5" and "Human 6" in Figure 1. We could not decide which paralog is the "parent" in the first round of whole genome duplication and which ones are the "parent" in the second round. Comparison of the synteny evidences of these 4 paralogs would not yield meaningful results. In condition 2, for the innermost duplication event, there are no synteny evidence for each of the paralogs that descended from that duplication event. Thus, in our model, no parent-daughter relationship could be inferred for this case. Tandem duplications produce identical adjacent gene duplicates. For example, a sequence of genes in a chromosome region A, B, C, D forms A, B1, B2, B3, C, D via tandem duplication. In our model, we cannot discern which of the "B1", "B2" and "B3" is the parent in the duplication event.

This analysis is a first attempt to combine the phylogenetic analysis and synteny block evidence to decipher a deep and comprehensive "Parent-Daughter" relationships in cases of multiple successive duplications among vertebrate paralogs. There are several limitations to this study. The phylogenetic tree structure was obtained from the PANTHER 10 database, a most widely used collection of gene families (Mi, et al. 2017). The accuracy of the "Parent-Daughter" inference is dependent on the correctly reconstructed duplication events at various ceancestors. The building of the phylogenetic trees has been a notoriously difficult problem; thus, the accuracy of the phylogenetic trees would be a major limitation of our study. The principle of PANTHER's reconstruction of the duplication events is through maximum parsimony. Large protein families are especially prone to potential errors. In addition, due to large scale gene losses and recombination, the "syntenic" evidence may be incomplete. In our previous analysis, we found only 55% of the duplication events in early vertebrates are from large segmental duplications and most of the duplication nodes at early vertebrates have fewer than 4 clades, indicating early gene losses after the whole genome duplications. Thus, we may have missed a large amount of the "Parent-Daughter" relationships in many gene families. In addition, we could not completely eliminate the possibility of large segmental duplications in ceancestors later than early vertebrates. Thus, it might be mistaken to treat the genes with synteny evidence as the "old" gene copy originated from the early vertebrate duplications.

## Conclusions:

Gene duplication is a major mechanism through which new genetic material is generated. Paralogs of a genome may come from multiple duplication events at different evolutionary periods, making deciphering the "Parent-Daughter" relationship of each duplication event a difficult task.

The hypothesized 2 rounds of whole genome duplications (2R WGDs) at early vertebrates provides a unique perspective for decoding the "Parent-Daughter" relationships. The whole genome duplications result in sequential homologous genes with conserved gene orders. Using 17 extant vertebrates, we extensively examined the within and between genomes synteny evidences by extracting genes that are descended from duplication at early vertebrates. Through additional phylogenetic analysis, we inferred duplication events at periods younger than early vertebrates, like Mammalia, or Primates. By examining descendants of these younger duplication events, we could identify the "Parent" gene from the later "daughter" gene as the "Parent" gene is from the 2R WGDs and should be located within synteny blocks.

Our study reveals that the "Parent" copy has significantly smaller accumulated amino acid changes compared with the "Daughter" copy, indicating that the "parent" copy is more resistant to mutations. When there are several rounds of duplications after early vertebrates, in 905 of 1481 (61.11%) cases, the "Parent" copy in an older duplication event continues to be the "Parent" of the younger duplication events which lead to "grand-daughters".

This study is a first tempt to reveal the deep history of gene duplications in vertebrates by combining phylogenetic analysis and examining syntenic evidences. It may shed light on the evolutionary history reconstruction of vertebrate evolution, and the mechanism of gene duplications.